\documentclass{article}
\usepackage{spconf,amsmath,graphicx}
\usepackage{booktabs}
\usepackage{xcolor}
\usepackage{amsmath}
\usepackage{amssymb}
\usepackage{booktabs}
\usepackage{arydshln}
\usepackage{tabularx}

\usepackage{pifont}
\newcommand{\cmark}{\textcolor{teal}{\ding{51}}}%
\newcommand{\xmark}{\textcolor{magenta}{\ding{55}}}%
\newcommand{\posmark}{\textcolor{teal}{P}}%
\newcommand{\negmark}{\textcolor{magenta}{N}}%
\usepackage{cite}

\usepackage[%
    colorlinks=true,
    pdfborder={0 0 0},
    linkcolor=green
]{hyperref}

\usepackage[capitalize,]{cleveref}

\crefname{section}{Sec.}{Secs.}
\Crefname{section}{Section}{Sections}
\Crefname{table}{Table}{Tables}
\crefname{table}{Tab.}{Tabs.}

\title{VoiceVector: Multimodal Enrolment Vectors for Speaker Separation}
\name{Akam Rahimi, Triantafyllos Afouras, Andrew Zisserman\thanks{This work is funded by the UK EPSRC AIMS CDT, the EPSRC Programme Grant VisualAI EP/T028572/1,
and a Google-DeepMind Graduate Scholarship.}}
\address{VGG, Department of Engineering Science, University of Oxford, UK}
\begin{document}
%
\maketitle
\begin{abstract}
We present a transformer-based architecture for voice separation of a target speaker from multiple other speakers and ambient noise. We achieve this by using two separate neural networks: (A) An enrolment network designed to craft speaker-specific embeddings, exploiting various combinations of audio and visual modalities; and (B) A separation network that accepts both the noisy signal and enrolment vectors as inputs, outputting the clean signal of the target speaker. The novelties are: (i) the enrolment vector can be produced from: audio only, audio-visual data (using lip movements) or visual data alone (using lip movements from silent video); and (ii) the flexibility in conditioning the separation on multiple positive and negative enrolment vectors. We compare with previous methods and obtain superior performance. 
\end{abstract}
\vspace{-5pt}
\section{Introduction}
\vspace{-5pt}
\label{sec:intro}
Speech and audio processing applications, such as hearing aids, voice-activated systems, and video conferencing, hinge on voice separation performance, especially in noisy, multi-speaker environments. The complexity of separating voices is accentuated when faced with diverse acoustic settings, ambient noise, and overlapping speech.

A notable set of methods represented by~\cite{wang2018voicefilter, SpeakerBeam8736286, ochiai19_interspeech}, addresses this challenge by conditioning the separation process on speaker embeddings that are derived from clean and noise-free audio. These embeddings capture the distinctive vocal characteristics of an individual, and while they have shown significant promise in controlled environments, their dependency on such clean audio can hinder their effectiveness in real-world, noise-rich scenarios where pre-recorded clean audio of the target speaker is not available.

On the other hand, a subset of methodologies, such as~\cite{Rahimi22, ephrat2018looking, Afouras20b}, harness visual cues, specifically synchronised lip movements, to enhance the voice separation task. These audio-visual methods, while effective, are not without limitations. Their optimal performance is often contingent on the uninterrupted availability of visual cues from the very recording that is being separated. This becomes a challenge, especially when visual data is occluded or missing.

\begin{figure*}[t]
  \centering
   \vspace{-32pt}
   \includegraphics[width=.99\linewidth]{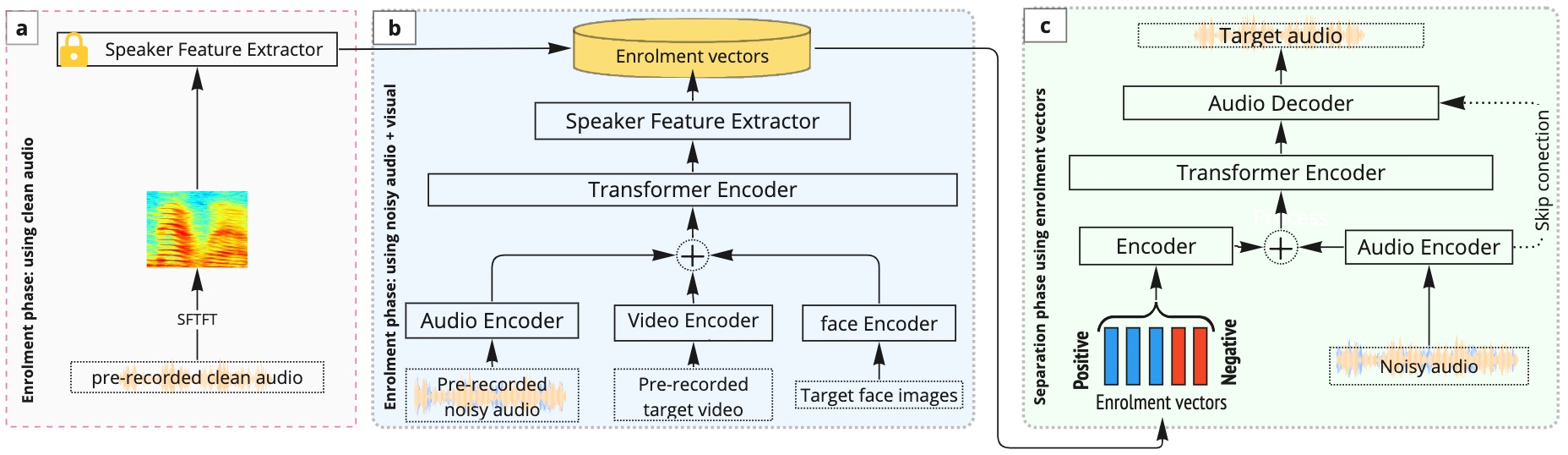}
   \vspace{-15pt}
   \caption{
The VoiceVector Architecture Overview: There are three networks for two tasks: `enrolment vector` generation networks ({\bf a,b}); and a speaker separation network ({\bf c}). In the enrolment phase, network {\bf a} is used to generate enrolment vectors from clean audio, while network {\bf b} accepts other data modalities, including: various combinations of lip-reading, facial images, or noisy audio paired with visual cues. In {\bf b} a combination of available modalities is concatenated in time dimension and passed to a transformer encoder which aggregates the concatenated input. In particular, during training, the speaker feature extractor in network {\bf b} is trained to mirror the output of network {\bf a} for separate recordings of the same speaker. In the separation phase {\bf c}, a U-Net structure augmented with a transformer, processes noisy audio conditioned on a set of positive and negative enrolment vectors generated from {\bf a} or {\bf b}. Using these vectors, the model separates the waveform of the target speaker.
   }
   \label{fig:architecturea}
   \label{fig:architectureb}
   \label{fig:architecturec}
   \vspace{-10pt}
\end{figure*}

In this paper, we present a novel method that marries the strengths of both the audio and visual conditioning paradigms while efficiently sidestepping their inherent challenges. Our approach is structured around two distinct phases:

\noindent \textbf{\textit{\small Enrolment Phase:}} In this phase, we harness auditory and/or visual information to craft `enrolment vectors'. These descriptors enable the separation of a target speaker's voice in noisy settings. Although our method can derive these vectors from clean audio (Fig.\ref{fig:architecturea}a), they can equally be crafted from noisy audio paired with visual cues or visual cues alone (Fig.\ref{fig:architectureb}b). Notably, unlike other audio-visual methods, ours doesn't mandate sourcing visual cues directly from the intended recording being separated. Furthermore, our approach uses not just positive examples from the target speaker, but also negative examples from other speakers, serving as a contrastive measure. This dual conditioning significantly improves the separation performance of the model.

\noindent \textbf{\textit{\small Separation Phase:}} For this stage, we only use auditory data (Fig.\ref{fig:architecturec}c). By design, we circumvent the reliance on visual cues during separation, rendering our system both computationally efficient and robust against the variability or absence of visual information, thus addressing the limitation of prior audio-visual techniques.

In summary, our contributions include:
(i) Introducing speaker-discriminative embeddings using a specialised enrolment network that is amenable to audio, audio-visual, and uniquely, video-only inputs.
(ii) Conditioning voice separation with both positive and negative enrolment vectors.

\vspace{-8pt}
\section{Related Work}
\vspace{-5pt}

Advances in target speaker separation have predominantly used two paradigms: multi-modal conditioning, and speaker-specific embeddings \cite{surveySpeakerSeparation}. Several methods represented by~\cite{Afouras18, Rahimi22, martel23_interspeech, Afouras20b,gabbay2018seeing,ephrat2018looking,Afouras19b}, use visual cues, mainly lip movements, to inform separation. While these methods require these cues throughout the separation, our separation phase operates entirely without them, offering a significant advantage. 

Turning to speaker-conditioned models, a common approach is to condition the separation on speaker embeddings generated from clean speech samples from target speakers~\cite{wang2018voicefilter, SpeakerBeam8736286, Snyder2018xvector, ochiai19_interspeech}. In a similar manner to our video-only (no audio) enrolment approach, the face images of the target speakers have also been used to separate the speakers~\cite{chung2020facefilter}. Previous research by \cite{xiao2019single} has used negative vectors for audio separation. In contrast to their approach, our model can generate the enrolment vectors from noisy audio and does not require distinct attention mechanisms for positive and negative vectors.

Attempts to separate the target speaker using only audio~\cite{luo2018tasnet, Hershey2016deepclustering}, are challenged by the permutation problem. While permutation-invariant training~\cite{Yu2017permutation, Kolbeak2017permutation} provides partial alleviation, the inference stage reintroduces the problem due to ambiguous channel assignment.

\vspace{-10pt}
\section{Method}
\vspace{-5pt}
\label{sec:pagestyle}
In this section, we describe our method, named `VoiceVector', for separating a target speaker from other simultaneous speakers and ambient noise. When presented with a noisy speech signal, the aim is to separate the component representing the distinct acoustic characteristics of a specified speaker, whilst concurrently filtering components such as competing voices or ambient noise. Refer to Fig.~\ref{fig:architecturea} for a graphical representation of the architecture. 
\vspace{-10pt}

\vspace{-2pt}
\subsection{Architecture}
\label{ssec:architecture}
\vspace{-5pt}
The overall architecture consists of two separate but interconnected networks, one to generate enrolment vectors and the other tasked with speaker separation.  

\vspace{-10pt}
\subsubsection{Speaker enrolment network.}
\label{sssec:speaker enrolment network}
\vspace{-5pt}
We have devised three distinctive configurations of the enrolment network, each tailored for specific input modalities:\vspace{4pt}

\noindent \textbf{\textit{\small Audio-Only Network (Fig~\ref{fig:architecturea}(a)):}} This configuration makes use of an off-the-shelf speaker feature extractor network named \text{\small ECAPA-TDNN} \cite{ecapatdnn}. It processes clean audio \( a_c\in \mathbb{R}^{t_{a}} \) represented in the spectrogram format, \text{\textit{\small S(f,t) = STFT}}($a_c$), and outputs speaker embeddings, \( S_{a_c} \in \mathbb{R}^{192} \). Here \text{\textit{\small STFT}} denotes the Short-Time Fourier Transform. \vspace{5pt}

\noindent \textbf{\textit{\small Audio-Visual Network (Fig~\ref{fig:architecturea}(b)):}} 
This network generates speaker embeddings from noisy audio by leveraging associated visual elements. This includes video footage that captures lip motions and/or still images of the target speaker.
The noisy audio waveforms \( a\in \mathbb{R}^{t_a} \) are encoded as \( E_a\in \mathbb{R}^{ t_a \times 768} \), and concurrently the video stream (lip motion) \( v\in \mathbb{R}^{3\times t_v \times H \times W} \) is encoded as video embeddings \( E_v\in\mathbb{R}^{t_v \times 512} \). Optionally, this can be further enriched by introducing static face images \( f\in \mathbb{R}^{3\times H \times W} \) of the target speaker encoded into image embeddings \( E_f\in \mathbb{R}^{128}\). $t_a$ and $t_v$ represent the temporal dimensions of audio and video, respectively. These encodings are passed through separate convolution layers to have the same channel dimensions and are concatenated to form a joint feature representation.

{
\vspace{-12pt}
\small
\[ F(E_a, E_v, E_f) = ( E_a; E_v; E_f) \in \mathbb{R}^{ ( t_a + t_v + 1)\times 768} \]
\vspace{-10pt}
\label{eq:concat}

}

The aggregated features are passed through a three-layer transformer encoder. The resulting encoded representation is then fed into a modified speaker feature extractor, producing the final enrolment vectors \( S_{avf}\in \mathbb{R}^{192} \). \vspace{5pt}

\noindent \textbf{\textit{\small Video-Only Network (Fig~\ref{fig:architecturea}(b)):}} %
In this case there is no audio input. The video encoder processes a silent video stream and outputs visual embeddings \(E_v \in \mathbb{R}^{t_v \times 512}\), while the face encoder transforms still images into face embeddings \(E_f \in \mathbb{R}^{128}\). The network can utilise either or both embeddings. When both are used, they are aligned by channel dimensions and concatenated. This combined data undergoes processing through three layers of transformer encoders, which is then fed into the modified speaker feature extractor to produce enrolment vectors \(S_{vf} \in \mathbb{R}^{192}\).


{
\small
\vspace{-15pt}
\[ S_{vf} = SpeakerExtractor (Transformer([E_v; E_f])) \]
}


\vspace{-16pt}
\subsubsection{Speaker separation network.}
\label{sssec:subsubhead} 
The second network shown in Fig~\ref{fig:architecturea}(c) is a VoiceFormer-inspired model~\cite{Rahimi22}, consisting of an audio encoder-decoder and a speaker embedding encoder. This network is fed with noisy audio waveforms and corresponding multiple positive and negative enrolment vectors. It uses a three-layer Transformer encoder to combine the audio and speaker encodings. Attention is increased for features that match the target speaker (positive embeddings), while downplaying features from non-target speakers (negative embeddings). Finally, skip connections between the audio encoder and decoder help retain both low- and high-level information, leading to improved performance. The output of this network is the waveform signals from the separated target speaker.

\vspace{-10pt}
\subsection{Training}
\label{ssec:training}
\vspace{-5pt}

\textbf{\textit{ Training objectives:}}
For the \textbf{\textit{\small{Speaker Enrolment Network}}} given
a dataset \(\mathcal{D}\) of tuples \((a, v, f)\), the network uses pre-recorded noisy audio waveforms \( a\in \mathbb{R}^{t_a} \), conditioned on lip motions alone \( v\in \mathbb{R}^{3\times t_v \times H \times W} \), face images alone \( f\in \mathbb{R}^{3\times H \times W} \) or a combination of both, to predict an enrolment vector \( S_p\in \mathbb{R}^{192} \). The objective is to minimise the \( L1 \) discrepancy between the predicted \( S_p \) and the speaker embedding \(S_{a_c}\) of the target speaker generated by the audio-only network shown (Fig~\ref{fig:architecturea}(a)). Crucially \(S_{a_c}\) is generated from a distinct recording of the target speaker.
 $\mathcal{L}_{\text{enrolment}} = \mathbb{E}_{(a, v, f) \in \mathcal{D}} \| S_{a_c} - S_p \|_1$.

For \textbf{\textit{\small{Speaker Separation Network}}}, given a dataset \(\mathcal{D}\) of tuples \((a_m, S_p, a_t)\), the network conditions the noisy audio mixture \(a_m\) on the predicted enrolment vectors \(S_p\) and aims to align the predicted audio \(\hat{a}_t\) with the clean target waveforms \(a_t\), formalised using the L1 loss: $\mathcal{L}_{\text{separate}} = \mathbb{E}_{(a_m, S_p, a_{t}) \in \mathcal{D}} \| a_{t} - \hat{a}_{t} \|_1$

\noindent \textbf{\textit{Training stages:}} Training is structured in a step-by-step process. Initially, as shown in Fig~1(a), clean audio is used to create target enrolment vectors. Next, as shown in Fig~1(b), the enrolment network is trained using knowledge distillation, so that its output aligns with the speaker feature extractor model in Fig~1(a). This network is further enhanced by training it with a mix of noisy audio with corresponding visual elements or just visual information. The video-only version capitalises on either the lip motion of the speaker or by integrating static full-face images of the speaker in tandem with the corresponding video clips. The subsequent step involves training the separation network, represented in Fig~1(c). Here, we use noisy audio that includes the voices of the target speaker, other speakers, and background noise. This audio is paired with a random combination of both positive and negative enrolment vectors that serve as conditioning signals, helping to emphasise the desired voice while suppressing others in the mix. Finally, we implemented an end-to-end fine-tuning phase, refining both the enrolment and separation networks.
\vspace{-10pt}
\subsection{Implementation details}
\vspace{-5pt}
The model is trained using Pytorch. Video data underwent facial cropping to the mouth region of the speaker with a uniform 25 FPS. The audio was monophonised via channel averaging and resampled to 16kHz.

The enrolment network employs a visual backbone from~\cite{Prajwal21}, integrating a \text{\small VTP} network with transformer units over a hybrid 3D/2D residual CNN. Face embeddings are produced via an off-the-shelf library based on the VGG-Face model~\cite{Parkhi15}. The audio encoder uses a sequence of 1D CNN layers, processing audio waveforms directly. The Transformer comprises 3 layers and 8 attention heads, with a model dimension of \(532\). We use 768-dimensional embeddings across modalities throughout, ensuring alignment with post-audio encoding channel dimensions. The intermediate outputs of the transformer within this network are channelled into an adapted \text{\small ECAPA-TDNN} model \cite{ecapatdnn}. Note that unlike traditional \text{\small ECAPA-TDNN} that consumes spectrograms, our modified version leverages features from the transformer's output.

For the speaker separation network, the architecture drew insights from the design in~\cite{Rahimi22}, but adapted to accommodate enrolment vectors as opposed to video or text. This module integrates an audio encoder-decoder with a dedicated encoder for enrolment vectors. The model ingests a varying number of positive and negative enrolment vectors, always ensuring at least one positive enrolment vector during training. A three-layer Transformer bottleneck fuse encoded noisy audio and enrolment vectors, attending to relevant audio features based on the positive and negative embeddings. 

Each enrolment vector is derived from a 4-second audio sample, processed using the enrolment network depicted in Fig. \ref{fig:architectureb} (b). During training, the separation network was exposed to a varied number of positive and negative vectors. 

\vspace{-8pt}
\section{Experiments}
\vspace{-5pt}
\label{sec:typestyle}

\subsection{Synthetic sequences}
\vspace{-5pt}
\label{ssec:Synthetic_sequences}
Consistent with previous research~\cite{Afouras18,ephrat2018looking}, we train and test our models using synthetic noisy samples, formed by combining waveforms from two separate speakers. For a diverse training set, we mix speech from multiple speakers and background noise from the DNS dataset, including traffic, wind, and other general noises. These 4-second audio mixtures are randomly cropped from extended tracks and undergo augmentations such as speed, pitch, and decibel adjustments to mimic varied acoustic scenarios. It is worth highlighting that despite our model's training and evaluation on synthesised audio mixtures, it remains adept at managing authentic noisy sequences, given the small domain gap between synthetic and real-world audio samples.

\vspace{-10pt}
\subsection{Datasets, training and evaluation protocol}
\label{ssec:Datasets}
\vspace{-5pt}
\textbf{\textit{ Data.}}
Our evaluation uses two well-known datasets:
LRS3~\cite{afouras2018lrs3ted} and Librispeech\cite{librispeech15}. LRS3 derived from public TEDx videos boasts a vast array of audio-visual clips covering diverse speaking styles and subjects. Traditionally favoured for lip reading and speech recognition, its comprehensive nature aligns perfectly with our objectives. Librispeech, a vast audio-only dataset sourced from public-domain audiobooks, is meticulously segmented, proving invaluable for tasks like automatic speech recognition and speaker identification.

Both the LRS3 and Librispeech datasets cover a diverse range of speakers of varying genders, accents, and speech patterns, offering a rich and varied testing environment. Crucially, the speakers in the test sets are not encountered during training. Synthetic test samples were crafted by randomly mixing two distinct speakers from the evaluation subsets of LRS3 and Librispeech, with the addition of ambient noise.


\noindent \textbf{\textit{Evaluation metrics.}}
 We use three standard metrics: (i) Signal-to-Distortion Ratio (SDR) to quantitatively measure the quality of the separated output; (ii) Short-Time Objective Intelligibility (STOI) quantifies signal intelligibility; and (iii) the Perceptual Evaluation of Speech Quality (PESQ) offers a perceptual score reflecting the listener's perception. 
 Together, these metrics offer a comprehensive evaluation of our method and allow comparison with existing benchmarks.

\vspace{-5pt}
\subsection{Results}
\vspace{-5pt}
\label{ssec:results}
In this section, we evaluate our proposed method. We first assess the separation performance under various combinations of positive and negative embeddings. Next, we evaluate the model based on enrolment vectors of various modalities. We conclude by benchmarking against the state-of-the-art in speaker separation and speech enhancement tasks.

\noindent \textbf{\textit{Positive and negative embeddings.}} The positive vectors are crafted from distinct recordings of the target speaker, ensuring they are different from the immediate audio being separated. On the other hand, the negative vectors are sourced from different recordings of the non-target speaker present in the mixed audio. Table \ref{tab:pos-neg} illustrates the effectiveness of introducing both positive and negative enrolment vectors. The results confirm that increasing the number of enrolment vectors enhances the model's separation accuracy.


\begin{table}[ht]
\vspace{-10pt}
    \caption{
    \textbf{Speaker Separation performance with a varied number of enrolment vectors.} 
    Results on the LRS3 test set, as the numbers of \posmark \, positive and \negmark \, negative enrolment vectors are varied. It is evident that increasing the number of positive and negative enrolment vectors enhances performance. Higher score is better.}
    \label{tab:pos-neg}
    \vspace{5pt}
    \setlength{\tabcolsep}{4pt}
    \centering
    
        \begin{tabularx}{\linewidth}{X cccc}
            \toprule
              & \small{1\posmark}  |  \small{0\negmark} \:& \small{1\posmark}  |  \small{1\negmark} \:& \small{3\posmark}  |  \small{2\negmark} \:& \small{3\posmark}  |  \small{3\negmark}  \\
            \midrule
            SDR$\uparrow$   & 13.8 & 14.0  & 14.4 & \textbf{14.5} \\
            \bottomrule
        \end{tabularx}
    
    \vspace{-5pt}
\end{table}

\noindent \textbf{\textit{Modalities comparison.}} 
Table \ref{tab:modalities} illustrates our model's effectiveness in the interplay between audio and visual modalities. When operated solely on clean audio, the model stands robust, but the introduction of visual cues, specifically lip motions, to the clean audio offers a marginal advantage. The model stands resilient in cases where we do not have access to any clean audio from the target speaker and the positive vectors are generated from a combination of noisy audio and visual cues. Interestingly, the reliance solely on lip motions, yields impressive results, highlighting the importance of visual cues. Furthermore, incorporating the facial data amplifies performance. This harmony between auditory and visual data shows the strength of our multi-modal approach in scenarios where one of the modalities might not be available.
\begin{table}[ht]
\vspace{-10pt}
    \caption{
    \textbf{Speaker Separation: Enrolment Vectors Across Modalities.} 
Symbols \cmark~ and \xmark~ indicate the modality's presence or absence in the enrolment network. A stands for audio and V for visual cues. We assess speaker separation using enrolment vectors generated from various modality combinations in the synthetic LRS3 test set. Using 3 positive and 2 negative vectors. Higher is better.
}
    \label{tab:modalities}
    \vspace{5pt}
    \setlength{\tabcolsep}{2pt}
    \centering
        \begin{tabular}{l cc ccc}
            \toprule
            Modality & A & V  & SDR$\uparrow$ & STOI$\uparrow$ & PESQ$\uparrow$  \\
            \midrule
            Clean audio-only    & \cmark & \xmark  & 14.4 & 91 & 2.52   \\
            Clean audio + lip motions   & \cmark & \cmark  & 14.5 & 91 & 2.55  \\
            Noisy audio only    & \cmark & \xmark  & 6.3 & 58 & 1.82  \\
            Noisy audio + lip motions    & \cmark & \cmark  & 13.7 & 88 & 2.45  \\
            Lip motions only   & \xmark & \cmark  & 11.1 & 77 & 2.25  \\
            Lip motions + face images   & \xmark & \cmark  & 12.0 & 80 & 2.35  \\
            Face images only  & \xmark & \cmark  & 10.8 & 79 & 2.20  \\
            \bottomrule
        \end{tabular}
    \vspace{-10pt}
\end{table}

\noindent \textbf{\textit{State-of-the-Art Comparison.}} 
Table~\ref{tab:baselines} compares the performance of our approach in speaker separation with preceding methods. For benchmarking, we consider speech separation conditioned on speaker embeddings from \cite{wang2018voicefilter} and a selection of recent audio-visual approaches \cite{Rahimi22,Afouras18,gao2021visualvoice}. VoiceVector consistently surpasses or performs on par with previous methods.
It is worth highlighting that, unlike the other audio-visual methods, our model does not access visual cues during separation, nor has it encountered synchronised visual cues from the mixture undergoing separation.

\begin{table}[htb]
\vspace{-10pt}
 \caption{\textbf{Comparison to the state-of-the-art on speaker separation.}
    Evaluation on the synthetic LRS3 and Librispeech test sets. Our best model outperforms the previous work.  A stands for audio and V for visual cues. \cmark~ and \xmark~ indicate the presence or absence of the modality respectively. $\dagger$Our model has not been trained on Librispeech, demonstrating generalisation across datasets.  Higher is better.
    }
    \label{tab:baselines}
    \vspace{5pt}
    \centering
    \setlength{\tabcolsep}{2pt}
    \begin{tabularx}{\linewidth}{l ccc ccc}
        \toprule
        Model & A & V & Testset & SDR$\uparrow$ & STOI$\uparrow$ & PESQ$\uparrow$\\
        \midrule
        Conversation ~\cite{Afouras18}   & \cmark & \cmark  & LRS3 & 11.5 & 87 & 2.18 \\
        VisualVoice ~\cite{gao2021visualvoice}   & \cmark & \cmark  & LRS3 &  11.9 & 90 & 2.46 \\
        VoiceFormer ~\cite{Rahimi22}  & \cmark & \cmark  & LRS3 & 14.4 & \textbf{92} & 2.42   \\
        Ours (VoiceVector)    & \cmark & \cmark  & LRS3 & \textbf{14.5} & 91 & \textbf{2.52} \\
        \midrule
        VoiceFilter ~\cite{wang2018voicefilter}  & \cmark & \xmark  & Librispeech & 12.6 & -- & --   \\
        Ours (VoiceVector)$^\dagger$  & \cmark & \xmark  & Librispeech & \textbf{13.1} & \textbf{89} & \textbf{2.12} \\
        \bottomrule
    \end{tabularx}
   
    \vspace{-5pt}
\end{table}
\vspace{-8pt}
\section{Discussion}
\vspace{-5pt}
\label{sec:majhead}
In summary, we introduce a two-phase approach to voice separation that leverages combinations of auditory and visual data to generate enrolment vectors, followed by a separation phase that exclusively relies on auditory data. This strategy ensures robustness against visual data variability in comparison to previous audio-visual separation approaches. The use of both positive and negative vectors exemplifies a nuanced angle to the separation task while improving performance.




\vfill\pagebreak
\bibliographystyle{IEEEbib}
\bibliography{refs}

\end{document}